\documentclass[preprint]{aastex}

\begin{document}

\title{A Near-Infrared Stellar Census of Blue Compact Dwarf Galaxies:
NICMOS\footnotemark[1] Detection of Red Giant Stars
in the Wolf-Rayet Galaxy Mrk~178}

\author{Regina E. Schulte-Ladbeck}
\affil{University of Pittsburgh, Pittsburgh, PA 15260, USA}
\email{rsl@phyast.pitt.edu}
\author{Ulrich Hopp}
\affil{Universit\"{a}tssternwarte M\"{u}nchen, M\"{u}nchen, FRG} 
\email{hopp@usm.uni-muenchen.de}
\author{Laura Greggio}
\affil{Osservatorio Astronomico di Bologna, Bologna, Italy, and
Universit\"{a}tssternwarte M\"{u}nchen, M\"{u}nchen, FRG}
\email{greggio@usm.uni-muenchen.de}
\author{Mary M. Crone}
\affil{Skidmore College, Saratoga Springs, NY 12866, USA}
\email{mcrone@skidmore.edu}

\footnotetext[1]{Based on observations made with the NASA/ESA Hubble 
Space Telescope obtained from the 
Space Telescope Science Institute, which is operated by the Association of 
Universities for Research in Astronomy, Inc., under NASA contract NAS 5-26555.}

\begin{abstract}

We observed the Blue Compact Dwarf/Wolf-Rayet galaxy Mrk~178 with the NICMOS
camera aboard HST. The galaxy is well resolved into individual stars
in the near-IR; photometry in J and H yields color-magnitude
diagrams containing 791 individual point sources. We discuss the
stellar content, drawing particular attention
to the intermediate age and/or old stars. 

Mrk~178 is only the second Blue Compact Dwarf galaxy
in which the red giant branch has been resolved, 
indicating stars with ages of at least 1-2~Gyr. 
This allows us to derive a distance of $\geq$4.2($\pm$0.5)~Mpc. 
The near-IR color-magnitude diagram also exhibits an abundance
of luminous, asymptotic giant branch stars. We find that this requires 
vigorous star formation several hundred Myr ago. Some candidate 
carbon stars are identified via their extreme near-IR color. 

We argue that Mrk~178 is fundamentally an old galaxy, based on 
the NICMOS detection of red giants underlying the blue, starburst core, 
and its extended, faint halo of redder color.

\end{abstract}


\section{Introduction}

Within the well-accepted framework of hierarchical collapse,
today's dwarf galaxies are considered to be the non-merged remnants of
small-scale primordial density fluctuations in the early Universe 
(White \& Rees 1978, Dekel \& Silk 1986,  Ikeuchi \& Norman 1987).
Dwarf galaxies are still the most common kind of galaxy in the Universe today,
and they may have been even more abundant in the recent past when they 
presumably were part of the Faint Blue Excess (e.g., Ellis 1997,
Marzke \& Da Costa 1997). What then, is the relationship between today's dwarf 
population and
the dwarf-like component of the faint-blue galaxies undergoing strong 
starbursts at z~$\approx$~0.5-1 (Koo et al. 1994, 1995, Ferguson \& Babul 1998,
Guzm\'{a}n et al. 1998)?
At the present epoch, we observe two morphologically distinct 
kinds of dwarf galaxies: the early-type dwarfs (dSph and dE) which, broadly
speaking, contain mainly old stars, little or no gas, and are generally situated in
galaxy clusters; and the late-type dwarfs (dIrr and BCD) which are gas rich and
actively forming stars. These are generally found in low-density environments with
some even in galaxy voids (Popescu, Hopp \& Rosa 1999).
How have dark-matter content, astration rate, gas cycling  and environment
produced this range of properties in the dwarf galaxy population we see
today (e.g. Dekel \& Silk 1986, Skillman \& Bender 1995, Ferrara \& Tolstoy 2000)?

It has been a long-standing question whether Blue Compact Dwarf (BCD) galaxies
are old galaxies which made their first generation of
stars at high redshift and are currently experiencing a burst of star formation,
or young galaxies which are forming their very first stars at the present 
epoch (Searle \& Sargent 1972, Searle, Sargent \& Bagnuolo 1973, Kunth,
Maurogordato \& Vigroux 1988, Thuan 1991, Izotov \& Thuan 1999). 
This question is prompted by the combination of low metal abundances (for 
ionized gas) and the high present-day star-formation rates in BCDs.  Abundance 
determinations have traditionally been used to gain insight into the histories 
of BCDs. Izotov \& Thuan (1999) recently
compared the chemical abundances of the ionized gas of 50 BCDs with theoretical stellar
nucleosynthetic yields, and concluded that most BCDs are young. 
Specifically, they argue that galaxies with metallicites
(oxygen abundances) Z~$\leq$~Z$_{\odot}$/20 are experiencing
their first starburst and are no older than $\approx$40~Myr to 100~Myr, so-called 
``primeval" galaxies. For those BCDs with Z~$\leq$~Z$_{\odot}$/5, they suggest
an upper age limit of 1-2~Gyr.

A new avenue to age-date BCDs has opened up with the advent of HST, which allows nearby
BCDs to be resolved into individual stars. Color-magnitude
diagrams (CMD) have been interpreted to indicate that some low metallicity
BCDs are neither primeval nor young. For example, the present starburst is not the first one
to occur in the famous I~Zw~18 (Aloisi, Tosi \& Greggio 1999), the most metal-poor BCD known 
($\approx$Z$_{\odot}$/50), and there is evidence that
VII~Zw~403 ($\approx$Z$_{\odot}$/20) may even contain an
ancient stellar population (e.g. Schulte-Ladbeck, 
Crone \& Hopp 1998, hereafter SCH98). 

The disagreement in the star-formation histories (SFH) derived from spectroscopy of the
ionized gas and photometry of the stellar content is a significant concern. CMDs can only be
obtained for very nearby galaxies, while emission-line spectroscopy can be used to
investigate star-forming galaxies at high redshift. It is therefore important to understand 
the data on local galaxies. 

CMD studies use stars as clocks to provide a measure of a galaxy's age. 
A very young galaxy may be recognized by its high-mass stars, which
live only for a few tens of Myr. 
When stars that are older than a few hundred Myr are detected --- 
 typically intermediate-mass stars on the early asympotic
gaint branch (E-AGB) or in the thermally pulsing AGB phase (TP-AGB) 
--- the galaxy is at least about 0.5~Gyr old. Note that in star clusters 
and early-type galaxies,  AGB stars of a much lower luminosity 
and mass are observed, possibly indicating intermediate ages of up to
10~Gyr. We discuss AGB stars
further below. In star-forming galaxies, many luminous AGB stars of comparatively high
mass are observed, and they swamp the contribution of the low-mass upper AGB
stars near the tip of the red giant branch (TRGB). The red-giant-branch 
(RGB) appears
after an evolutionary time scale of about 1~Gyr (Sweigart, Greggio \& Renzini 1990, 
Bressan, Chiosi \& Fagotto 1994). This result is robust and does not depend much
on the initial metallicity of the population.
Hence, when a galaxy exhibits a well-populated red giant branch, 
it is at least 1-2~Gyr old. The unambiguous detection of stars much older 
than 1-2~Gyr is a problem, 
because BCDs are situated at large enough distances to prevent direct
access to old main-sequence turnoffs, or horizontal-branch stars, with today's
telescopes. We refer to this limitation as the ``1-2~Gyr frontier". In any event,
an old galaxy would be one that contains stars with ages in excess of 10~Gyr.
The detection of a well defined TRGB is regarded as a necessary, but not a
sufficient condition to establish the presence of old stars (Da Costa 1994).

In this paper we report on our NIC2 observations of the BCD Mrk~178 = UGC~6541.
The advantage of using NICMOS to search for intermediate-mass/old stars in
this Wolf-Rayet (W-R) galaxy is that the spectral energy distribution of such stars
peaks in the near-IR. 
In Figure~1, we display the Digitized Sky Survey (DSS) 103a-E image of Mrk~178.
This image suggests that Mrk~178 belongs to the type ``iE" BCDs in the
classification of Loose \& Thuan (1986). In iE BCDs, the bright 
blue star-forming regions are situated 
near the center of an elliptical, faint red background sheet. The contour maps
of multi-filter CCD images by Gonz\'{a}lez-Riestra, Rego \& Zamorano (1988)
support this classification as well.
Mrk~178 was recently resolved into its brightest stars from the ground, resulting
in a distance modulus of 27.73, or a distance of 3.5~Mpc (Georgiev, Karachentsev 
\& Tikhonov 1997). The present-day metallicity derived from emission-line spectroscopy of
the ionized gas is 12+log(O/H) = $7.950^{+0.020}_{-0.021}$ (Kobulnicky \& Skillman 1996),
or $\approx$~Z$_\odot$/10. Mrk~178 is located at the peak of
the metallicity distribution observed for BCDs (compared with the sample  
of Izotov \& Thuan 1999). There is evidence for active star-formation over the 
last 10~Myr from the presence of emission lines in the spectra. Observation of 
the W-R bump in the spectrum of the younger of 
two centrally located H-II regions (Gonz\'{a}lez-Riestra, Rego \& Zamorano 1988)
suggests that this activity is on-going. The galaxy has an H-I detection in the survey
of Thuan \& Martin (1981) which, at the distance of 4.2~Mpc which we
derive in this paper,
translates into an H-I mass of about 1.4x10$^7$~M$_\odot$. By all accounts then, 
Mrk~178 appears to be a typical representative of the BCD class.

\section{Observations and reductions}

We observed Mrk~178 on 1998 September 27 as part of GO program
7859. The NIC2 camera, which has a field of view of 19\farcs2x19\farcs2, 
was centered at (J2000) R.A. 11:33:28.82 and Dec. 49:14:12.4 on the SE knot, the
brighter of two prominent star-forming regions (cf. Gonz\'{a}lez-Riestra, 
Rego \& Zamorano 1988). The galaxy UGC~6538 is nearby on the sky, but while 
Mrk~178 has a heliocentric velocity of 249($\pm$4)~km~s$^{-1}$,
UGC~6538 has a much larger recession velocity of 3180~km~s$^{-1}$ (see NED),
and is physically unrelated. The classification of Mrk~178 as a ``galaxy pair",
which is sometimes found in the literature, is not justified. Instead,
Mrk~178 appears to be an isolated star-forming dwarf.

The data were obtained in the F110W and the F160W filters, which are similar to
the ground-based J and H bands. Information about the observations can be gleaned
directly from the STScI WWW pages linked to program ID 7859; this provides
a detailed observing log.
The total integration time was about 5568~s in
F110W, and 2880~s in F160W. Specifically, the data set in F110W consists
of six individual exposures.
In the first set of three, the exposure times were 895.956~s per exposure, in set
two (less acquisition overhead in second orbit), they were 
959.953~s per exposure. The F160W observations consist
of one set of three exposure, each 959.953~s in duration. Between
individual exposures of a set, the camera was dithered by 1" in the X
direction using the canned, XSTRIP-DITH pattern. All exposures were read 
out in MULTIACCUM mode; cosmic rays were rejected.

The color image is displayed as Figure~2, and
shows how well Mrk~178 resolves into single stars with NIC2. 
The data were reduced following the steps outlined
in SHGC99. Our major concern was to reproduce as closely
as possible the reduction of the VII~Zw~403 data, from which
we derived our distance calibration. The calibration files
and the pedestal-removal software used reflect the status of knowledge
as of mid 1999. We used a set of temperature-dependent dark current files 
for the reduction of all data. These dark reference files correct 
for shading, which is a temperature-dependent component of detector bias.
We also investigated all of the data in our program for cosmic-ray persistence, and
pedestal problems. We used M. Dickinson's software for pedestal removal.
Even after the above re-reduction steps were carried out, the data had a 
spatially non-uniform background, a feature that poses a problem for
DAOPHOT (Stetson, Davis, \& Crabtree 1990). Therefore, the star-subtracted 
background was smoothed as described in SHGC99. We used the combined, mosaicked images 
that resulted from calnicb for the photometry, since we found that there was 
no significant difference to the DAOPHOT PSF photometry using the drizzle procedure. 
We checked the resulting photometry by eye, to make sure that we did not 
pick up spurious sources due to blending.

The conversion of PSF photometry to absolute photometry
in the HST Vegamag system was carried out following the prescriptions
given in the NICMOS Photometric Calibration CookBook. We determined the 
countrates of 9 (F110W) and 7 (F160W) well isolated stars (from the list
of stars used to define the PSF for DAOPHOT). These
countrates were measured in the aperture of 0\farcs5 to which the
NICMOS photometric calibration is referred. The sky-background was measured in
a ring between 0\farcs65 and 6\farcs0 from the respective star. 
To convert the countrates to fluxes, we used the following photometric keywords: 
for F110W, PHOTFNU = 1.823290 $\times$ 10$^{-6}$ Jy~sec~DN$^{-1}$, 
for F160W, PHOTFNU = 2.070057 $\times$ 10$^{-6}$ Jy~sec~DN$^{-1}$. 
We applied the correction factor of 1.15 in flux (or 0.1517 in mags) 
to the nominal infinite aperture. We adopted fluxes of 1898~Jy in F110W, 
and 1113~Jy in F160W, for the zeropoints of the HST Vegamag system. 
We then compared the photometry of the PSF stars in this system to their PSF photometry
in our internal, instrumental system. The resulting transformations were applied to 
the DAOPHOT star lists.

There are 1557 individual
point sources detected in F110W, and 1087 sources detected in F160W. Figure~3
shows the error distributions for the photometry, and Figure~4 shows the
results of tests which
investigate the incompleteness of our photometry. The results indicate that the 
limiting magnitude is about 26.5 in F110W and 
25.5 in F160W, and that the data are better than 90\% complete to a magnitude of
about 24.0 in F110W and 23.3 in F160W.

The completeness tests used the DAOPHOT/ADDSTAR routine. 
Tests were done in one-magnitude bins, ranging from $\approx$6 mag above
the detection limit to almost the detection limit. In every simulation,
100 artificial stars were added, and then the manipulated 
frames were measured in the same way
as the original ones. For every magnitude bin in every band, 10
simulations were carried out. The final star list of each test was
cross-correlated with the input list of artifical stars, and from the
difference, the completeness factors and errors shown in Fig.~4 
were calculated.

In the faintest magnitude bin, very few stars (5-13) were recovered. There is
indeed a non-zero chance that a faint artificial star is put on top of a
noise peak (or on another faint, but real object), which make it detectable,
while all others are not. Looking at the distribution of
input-magnitude versus output-magnitude lists of the tests, one gets a more-or-less
Gaussian error distribution for the brighter objects, which widens more
and more going fainter, as expected. When approaching the detection limit,
this distribution has a highly non-symmetric appearance: 
more and more test stars come back brighter than their input flux. In the
faintest bin, all recovered stars have brighter magnitudes than their
input, pointing clearly to the above effect. Naturally, this also happens 
in the real data. Further, at the very detection limit, unusually powerful 
noise peaks of the right shape may enter the target list. Therefore, all 
sources in the faintest magnitude bin above the detection limit are highly 
questionable. This must be considered in the discussion of the data.
As the total number of objects in the bins affected by blending is small, 
they should have little influence on the interpretation of the data, 
or the CMD simulations.

We merged the two star lists requiring a positional source coincidence of better
than 1.5~pixels or $\approx$0\farcs1 (determined from cross-correlation
experiments). There are 791 objects with coincident positions
in F110W and F160W. Figure~5 shows CMDs constructed from these data.

\section{Results}

The foreground galactic extinction towards Mrk~178 is negligible: A$_B$=0.000
from H-I maps (Burstein \& Heiles 1984); A$_B$=0.077, which corresponds to
A$_J$=0.016 and A$_H$=0.010 from IRAS maps (Schlegel et al. 1998, and see NED). 
Having only two bands available to us,
we cannot investigate the internal extinction using a two-color diagram,
but we assume the effect is small 
 in the near-IR.
Because the field of the NIC2 camera is so small and because Mrk~178 is
situated at high galactic latitude (63\fdg3), we expect the contamination
of the CMDs by galactic foreground stars to be very small. Thus, it is relatively 
straightforward to interpret the CMDs.

The CMDs of Fig.~5 are characterized by a strong red plume, and a comparatively
weak blue plume. Most of the stars in these CMDs appear at red colors and faint
magnitudes;  {\it this is the red tangle which contains the RGB}. Depending on
the specific star-formation history (SFH), 
some combination of AGB stars and even blue-loop (BL) stars can also contribute
to this feature. The solid line of Fig.~5 marks the magnitude at which we identify
the TRGB (see below). Above the TRGB, we expect to see AGB stars and red supergiants
(RSG). In Mrk~178, there is a considerable number of red stars above the RGB, which 
we suspect are AGB stars.
Above the red tangle, the linear feature extending to bright
magnitudes is presumably dominated by RSGs, whereas the reddest of the bright
stars, which form a fan-shaped grouping in the CMDs, are likely to be mainly AGB stars.
The blue plume can include main-sequence (MS), BL, and blue-supergiant
(BSG) stars.

\subsection{The distance}

The core helium flash of low mass stars occurs at a constant bolometric magnitude
over a wide range of stellar masses and metallicities. The TRGB has therefore been successfully
calibrated as a distance indicator for galaxies (Lee, Freedman \& Madore 1983). 
In I-band luminosity functions, the TRGB is found as a sharp rise in stellar frequency 
towards fainter magnitudes.
The bolometric correction has been determined empirically from globular cluster 
observations and the V-I color is well calibrated as a function of metallicity. 
After applying the bolometric correction, the distance modulus of a galaxy 
may be derived. This is the basis of the I-band TRGB method.

While RGB stars are much brighter in J and H than in I and V, affording
detectability of the TRGB to larger distances, there are several disadvantages in using the
TRGB in J and H as a distance indicator. In SHGC99, we investigated the TRGB in J
and H using both stellar evolution models and globular cluster observations. 
In stellar models, the TRGB in both J and H depends on metallicity; at low metallicity
the absolute magnitudes at the TRGB are fainter than at high metallicity. This is
illustrated in Fig.~8 of SHGC99. According to the Padova stellar models, the variation of
the absolute H-band magnitude with metallicity is small for low metallicities, indicating
that the H-band might be the better of the two bands to use for determining distances.  
In addition, there is a large uncertainty of the ground calibration of the J band, which 
prompted us to discard the TRGB in J in the ground system as a useful
distance indicator. 

In what follows we discuss the TRGB based both on the F110W, F160W data and the transformed,
H-band data. We also consider that we have no prior knowledge of the metallicity
of the RGB stars in Mrk~178, and give both our preferred ``short" distance based
on the assumption of a low-metallicity RGB, and a ``long" distance allowing the
H-band metallicity to be as high as solar. 

In order to determine the distance of Mrk~178, we need to measure the brightness
of the TRGB. This is usually accomplished by finding the steepest gradient in the
luminosity function of the stars in the red plume, e.g., by applying an edge-detecting, 
Sobel filter. In the case of Mrk~178 the situation is complicated by the fact that this 
galaxy has a prominant population of luminous AGB stars, which overlaps the RGB population 
near the TRGB. The net effect is that the contrast in stellar number counts at the TRGB 
is less pronounced, and the TRGB itself is harder to pin down. 
We combined information from luminosity functions with our evaluation of the CMDs, 
to settle on a best estimate for the magnitude at the TRGB in Mrk~178.
In Figure~6, we show the luminosity functions for red stars in the color interval
0.75~$<$ (F110W-F160W) $<$~1.5, in magnitude bins of 0.1. 
In the F110W filter, there is a sharp
rise at a magnitude of 24, which we identify with the TRGB. 
It is not entirely clear where to identify the TRGB in F160W; a rise is seen
across several bins.  Luckily, the CMDs can
be used to help understand the onset of the TRGB. While the feature which we consider
to be the TRGB remains constant in magnitude over a wide color range
in the [(F110W-F160W), F110W] CMD, it
is slanted in the [(F110W-F160W), F160W] CMD.  
The plateau in the [(F110W-F160W), F110W] CMD of Fig.~5 translates into a sharp jump 
in the luminosity function in the F110W filter in Fig.~6, while the slanted TRGB
in the [(F110W-F160W), F160W] CMD translates into a more gradual rise in the F160W 
luminosity function. This is reflected by our error estimates.
Using both the luminosity functions and the CMDs for guidance,
we determine the following magnitudes at the TRGB:

\begin{center}
m$_{F110W_o, TRGB}$~=~24.00~$\pm$~0.05  and\\
\smallskip
m$_{F160W_o, TRGB}$~=~22.7~$\pm$~0.2. \\
\end{center}

The errors reflect our estimates as to how well we think that we can
locate the TRGB in the data. We notice that we find the TRGB at magnitudes
where the data are still 90\% complete or better.
The distance modulus of Mrk~178 may be derived using the
calibration which we established for VII~Zw~403 in SHGC99

\begin{center}
M$_{F110W_o, TRGB}$~=~--4.28~$\pm$~0.10~$\pm$~0.18  and\\
\smallskip
M$_{F160W_o, TRGB}$~=~--5.43~$\pm$~0.10~$\pm$~0.18  \\
\end{center}

\noindent where the first error is the statistical error and is dominated by
how well we can determine the location of the TRGB in VII~Zw~403, and the
second one is the systematic error primarily due to the RR~Lyrae distance
calibration of the TRGB (see SCH98). 
In what follows, we will not give
the systematic error again explicitly; the errors quoted are those
that refer to our statistical errors. 
Applying this calibration directly provides the following distance moduli 
for Mrk~178 using F110W and F160W, respectively: 
28.28($\pm$~0.11) and 28.13($\pm$~0.22), 
corresponding to distances  of 4.5($\pm$~0.2)~Mpc and 4.2$^{+0.5}_{-0.4}$~Mpc.    
Application of the above calibration
assumes the RGB stars of Mrk~178 have the same metallicity as those
in  VII~Zw~403 ([Fe/H]=--1.92). 

We derive a more general calibration over a wider range of metallicities 
by comparing the
H-band magnitude of the TRGB of VII~Zw~403 with that of GCs and
stellar evolution models (cf. SHGC99). The TRGB
in H is a constant as a function of metallicity for a wide range of low
metallicities

\begin{center}
 M$_{H_o, TRGB}$~=~--5.5 $\pm$ 0.1\\
 for --2.3$<$[Fe/H]$<$--1.5.\\
\end{center}

We employ the equations given in SHGC99 to transform from F110W, F160W to ground-based
J and H. Figure~7 shows the transformed CMDs, on the same scale as those of Fig.~5
for comparison. The transformation has the effect of shifting the red plume closer
to the blue plume. The transformed H magnitude at the TRGB is

\begin{center}
m$_{H_o, TRGB}$~=~22.6~$\pm$~0.1. \\
\end{center}

\noindent We obtain an H-band
based distance modulus of 28.1($\pm$~0.26), corresponding to a distance 
of 4.2($\pm$~0.5)~Mpc.

If the RGB population of Mrk~178 is more metal-rich, the TRGB indicates
a larger distance modulus.  If the RGB stars 
of Mrk~178 have solar metallicity, then M$_{H_o, TRGB}$ is --6.2 in 
the stellar evolution models of the Padova group (e.g. Fagotto et al. 1994, 
Bertelli et al. 1994). 
This leads to a distance modulus of 28.8, or a distance of about 5.8 Mpc.
It is not entirely clear how well the metallicity dependence of the models 
compares with observations. In Figure~11 of SHGC99, we compare the Padova models
with observations of globular clusters (GC). While there is good agreement for the
lowest metallicities, the available data do not help to elucidate what M$_{H_o, TRGB}$
is at higher metallicities; there is too much scatter.

Our minimum distance estimated from the TRGB in the near-IR using the low-metallicity assumption,
4.2~Mpc, is slightly larger than the distance derived by Georgiev, Karachentsev \& Tikhonov (1997), 
3.5~Mpc. This is not surprising, mostly because the supergiant distance
method suffers from large uncertainties (see Schulte-Ladbeck \& Hopp 1998). 
Our errors are primarily due to the large component of AGB stars near the RGB tip. This problem could have been
avoided in principle, by pointing the NIC2 into the background sheet of Mrk~178. However, 
we decided to observe on the star-forming regions to minimize the risk --- if there were
no RGB or if the galaxy were more distant, we might have ended up not detecting any
stars in Mrk~178. We studied luminosity functions using only spatial regions near the rim
of the NIC2 chip, and there was no significant difference to where we located the TRGB
in that data compared to using data from the center of the chip. 

In the subsequent sections of this paper, we compare the Mrk~178 data with theoretical tracks
and synthetic CMDs. We point out that any ``high-metallicity" assumption 
for the RGB corresponds to a ``long" distance.  From our previous work, we
have a slight preference for the ``low-metallicity" assumption.
However, without better knowledge of the metallicity
of the RGB stars in Mrk~178, the ``low-metallicity" or ``short" distance used here must be
regarded as a lower limit to the true distance of Mrk~178.

\subsection{Distance-dependent global parameters}

We rescaled the global parameters to a distance of 4.2~Mpc (m-M=28.1).
At this distance, the H-I flux measured by Thuan \& Martin (1981) corresponds to
an H-I mass of 1.4x10$^7$~M$_\odot$. Gonz\'{a}lez-Riestra, Rego \& Zamorano (1988) measured
the H$_\alpha$ fluxes of the two prominent H-II regions. Since these are 
spectroscopically derived fluxes, they probably missed some flux outside
of the measurement aperture; the sum of the two fluxes will only give a lower limit on the 
star-formation rate (SFR) of Mrk~178. We derive L(H$_\alpha$) 
$>$~3.9x10$^{38}$~ergs~s$^{-1}$, and, following Hunter \& Gallagher (1986) who adopt
the Salpeter IMF from 0.1 to 100~M$_\odot$, the SFR is $>$~0.003~M$_\odot$~yr$^{-1}$.
New imaging data in H$_\alpha$ (Hunter 2000) suggest a value of 
$>$~0.008~M$_\odot$~yr$^{-1}$ (assuming zero extinction).
The diameter of Mrk~178 (D$_{25}$ from the RC3) 
is 1\farcm23. This corresponds to a linear size of 1.5~kpc. The total galaxy colors 
in the RC3 are listed as B-V=0.35, U-B=--0.3. These are rather typical 
colors compared to the colors of the dIrr/BCD sample considered
by Schulte-Ladbeck \& Hopp (1998).
The total apparent blue magnitude from the RC3, 14.50, translates into a total 
absolute blue magnitude of --13.60.
This corresponds to a blue luminosity of about 4x10$^7$~L$_\odot$ 
(for M$_{B,\odot}$=5.41). The ratio of the
H-I mass to the blue luminosity (in solar units) is 0.4. This value is typical 
compared with the compilation of dwarf-galaxy M$_{HI}$/L$_B$ ratios given in Huchtmeier, 
Hopp \& Kuhn (1997).

\subsection{The stellar content}

In Figure~8, we display [(J-H)$_o$, H$_o$] CMDs of Mrk~178 with three 
values of the DAOPHOT errors in both J and H applied as a selection criterion.
In comparison with Fig.~7, this shows that many of the faint and very red
stars have large photometric errors. In addition, some might well
be blends, as sometimes happens near the detection 
limits of crowded-field photometry. What can also be seen is that
even when only stars with small photometric errors are selected,
the salient features of the CMD remain. In particular, the blue
stars are real detections even in the near-IR. More importantly for
deriving the SFH of the galaxy, the detection of AGB and RGB stars is 
a secure result. 

In Figure~9, we overplot onto the [(J-H)$_o$, H$_o$] CMD  
stellar evolutionary tracks drawn 
from the same database which we used in SHGC99. 
Briefly, these tracks are based on the Padova library of stellar
tracks (e.g. Fagotto et al. 1994, Bertelli et al. 1994) and the stellar 
atmospheres of Bessell, Castelli \& Plez (1998).
We employ in our discussion two sets of tracks with metallicities of
Z=0.0004 and Z=0.004 (Z$_\odot$/50 and Z$_\odot$/5, respectively.)
The present-day ionized gas abundance of Mrk~178 (Z$_\odot$/10) is bracketed
by these two, and so these tracks are a reasonable approximation for any
``high-metallicity" and ``low-metallicity" stellar population
which we might expect to find.  Note that in the low-Z case we assume 
the short distance modulus and for the high-Z case we assume a long
distance modulus of 28.8. The tracks do not extend to the extremely 
red colors observed for some of the asymptotic
giants; this is due to the truncation at the first thermal pulse 
of the electronically available Padova tracks (Fagotto et al. 1994). 

When converting the Padova tracks to the observational
plane, we employed a different set of stellar atmospheres than the one used by the 
Padova group. Our low-metallicity tracks (Z=0.0004) have M$_{H_o, TRGB}$
in good agreement with the original Padova tracks. Both sets of tracks with
different atmospheres yield the same
distance. For the Z=0.004 tracks, our M$_{H_o, TRGB}$
is about 0.1 magnitudes brighter than the Padova one. This is not in conflict with the 
observations (see Fig.~11 of SHGC99). 
We now trace through the evidence for both
recent star formation and an older underlying population using the CMD
of Fig.~9.

There are three indicators of present-day star-forming activity in
Mrk~178. First, spectroscopic evidence exists for the presence
of W-R stars (Gonz\'{a}lez-Riestra, Rego \& Zamorano 1988). 
Star-formation in Mrk~178 must thus have been active within the last
few Myr. Given the present-day metallicity of Mrk~178, stars with 
very high masses,  roughly 50~M$_\odot$ and higher,
are expected to be above the cut-off mass for the W-R channel 
(see Maeder \& Conti 1994). Second, the H-II regions indicate the presence 
of ionizing, young and massive stars with ages of up to 10~Myr. 
Third, the blue plume of the CMD presumably contains
massive MS stars and BSGs.  The blue plume is fairly weak in Mrk~178. 
However, a comparison of the observations with the stellar-evolution tracks 
indicates that this is because the observations barely contain the top of the MS. The
MS turnoff of the 9~M$_\odot$ track, for example, is already well below our 
detection limit. 

Due to the sparsely populated blue plume of Mrk~178, we cannot
tell about internal extinction within the H-II regions from these data.
If anything, the effect appears to be small as the data scatter about the 
theoretically expected location of the upper MS.

The red plume is quite strong in Mrk~178. It can be split into regions  
above and below the TRGB. The luminous portion of the red plume can in 
principle be composed of RSG stars and AGB stars. Whereas optical colors, 
or optical---near-IR colors (see SHGC99), separate well many AGB stars from 
the RSG stars, these objects tend to overlap more closely in (J-H)$_o$. 
This observation is consistent with stellar-evolution models;
comparing the two sets of tracks with the data indicates an
age-metallicity 
degeneracy. We could interpret the grouping of high-luminosity stars 
at M$_{H_o}$ of about -8 and colors of (J-H)$_o$~$\approx$~0.7 either
as intermediate-mass stars, or as high-mass stars. Thus the possible age-range
spans from a few ten, to a few hundred Myr.

We can more clearly separate the AGB from the RSG stars observationally when 
they are redder than (J-H)$_o$ of about 1.0, in the regime of the 
thermally pulsing, or TP-AGB phase. Indeed we see in Figs.~8 and 9  a few stars
which have small errors, are bright, and are as red as (J-H)$_o$~$\approx$~1.5.
TP-AGB stars in principle probe ages of a few hundred~Myr to a few Gyr. 

The Z=0.004 tracks provide a fairly good
description of the luminous stellar content of Mrk~178. The Z=0.004 
low-mass tracks are, however, too red at the TRGB. The lifetime
of the 1~M$_\odot$ star of this metallicity is 8.5~Gyr. 
Tracks of lower masses, and older ages, fall to the right
of this track. This would have the effect of producing a 
red tangle which is too red to match the observations. 
This is illustrated by the synthetic CMDs discussed below. 
The low-mass tracks at 
low-metallicity (Z=0.0004), on the other hand,  on which
we find stars as old as the Hubble time, fall near the center of the distribution of
stars in the red tangle.
The fainter and bluer portion of the red tangle could
contain intermediate mass stars (with ages as young as 300 Myr).
Toward high masses 
(and young ages), the tracks are bluer and the TRGBs fainter. 
This may be what gives the red tangle its characteristic, slanted tip shape 
towards blue colors in [(J-H)$_o$, H$_o$] CMDs. 
The red tangle may thus in principle comprise a mix 
of old and intermediate-mass stars of low metallicity, as well as intermediate-mass
stars of higher metallicity. While the red tangle could contain stars
as old as the Hubble time, we cannot clearly pinpoint any ancient
stars in these data.

\section{Discussion}

We are interested in finding out whether Mrk~178 is fundamentally  young 
or old. To this end, we provide a variety of arguments that
help us glean its SFH. There is no doubt that stars of ages between 1-2~Gyr
are present, and hence Mrk~178 cannot be a primeval galaxy. Without stars in the
1-2~Gyr range of ages, we would not see an RGB or get a defined TRGB.
Whether or not Mrk~178 is an ancient galaxy is another problem entirely, because
the 1-2~Gyr frontier prevents us from 
directly identifying the old stars. However, we build a series of
arguments which indicate that the presence of ancient stars in Mrk~178
is probable. These include the presence of carbon stars, the galaxy's morphology, 
and constraints on the total mass in stars derived from synthetic CMDs.

\subsection{Comparison with other galaxies}

We recently presented and discussed CMDs of VII~Zw~403 (SCH98, SHGC99, Schulte-Ladbeck et al.
1999, hereafter SHCG99). HST/WFPC2 
observations of this very nearby BCD allowed us to demonstrate the presence of
intermediate-age AGB stars, and of a radial stellar population gradient from
the young ``core", to the old ``halo". The extended, old halo exhibits no young stars and
a narrow RGB which is
consistent with the presence of low-metallicity, low-mass, and consequently,
truly old stars. We argue that this galaxy is at least several~Gyr old, 
and possibly older than 10~Gyr. 

Optical CMDs are relatively well-understood.  To take advantage of 
this knowledge in interpreting near-IR CMDs, in SHGC99 we cross-identified 
the stars found in the optical observations of 
VII~Zw~403  with those found in the near-IR. 
The results are illustrated in Fig.~10, where we color-code stars
based on their location in the [(V-I)$_o$, I$_o$] CMD.  We color the stars 
in the blue plume in blue, the stars in
the red plume above the TRGB in magenta, and stars in the red tangle below 
the TRGB in red.  We show
stars in the red tail in black; this region of
the [(V-I)$_o$, I$_o$] CMD above the TRGB and at very red colors 
is populated by AGB stars and nothing else.  It is 
obvious that RSG and AGB stars overlap
in the red plume of a [(J-H)$_o$, H$_o$] CMD and that AGB stars are difficult
to separate from RSG stars based on near-IR color. Using near-IR colors alone, on the other
hand, we were able to identify a few additional, very red (in J-H) stars which 
were not seen in the optical. In Fig.~10, we coded the optically-identified AGB stars with
black dots. We coded all stars found in the [(J-H)$_o$, H$_o$] CMD of VII~Zw~403
which have a color $>$1.0 and are above the TRGB with open circles. This illustrates
the component of AGB stars found only in near-IR observations.

Mrk~178 is at a very similar distance as VII~Zw~403.
We transfer the above classification scheme to Mrk~178 in the following way.
All stars with color $<$~0.5 are shown in blue and are considered to be
blue plume. Stars with colors 0.5 $<$~(J-H)$_o$ $<$~0.85 and H magnitude
above the TRGB are shown as belonging to the red plume. Stars with
colors  above the TRGB are considered AGB stars, but only
the ``IR-detected ones in VII~Zw~403" (colors $>$~1) are emphasized with black dots, 
and the remaining
objects are coded part of the red tangle.

Comparing the luminous stellar content of VII~Zw~403 and Mrk~178, 
the most striking differences are the absence of very luminous RSGs 
and the presence of a very luminous BSG in Mrk~178. Small-number statistics 
could account for this. 
Another region where the two galaxies appear to differ is
in the region between the two plumes. 
This could be an age effect --- VII~Zw~403 exhibits no
W-R feature (Izotov 1998, private communication; Martin 1998, private
communication) and could already be in a post-starburst phase. It could also
be a metallicity effect --- VII~Zw~403 has a lower present-day
gas metallicity. Assuming the stars are also of lower metallicity, 
we expect to see more blueward-extended blue loops in VII~Zw~403.

There are a few very red stars at (J-H)$_o$~$\approx$~1.5 in both VII~Zw~403 and
Mrk~178. We verified that the red objects in Mrk~178 are
indeed point-source-like in appearance. These stars can only be some kind 
of long-period variable or
very possibly, carbon stars. The near-IR observations of Feast et al. (1982) 
indicate that Mira stars tend to occur with (J-H)$_o$ colors $\approx$~1.0. 
Colors redder than ~1~mag, of up to about 1.2~mag, 
are preferrentially seen in variable and non-variable stars of the S-type.  The 
reddest colors, up to about 1.7~mag, are reached only by carbon stars.
Spectroscopic follow-up observations are needed to unambigously classify
the very reddest stars in Mrk~178, but color selection based on near-IR colors
has been one of the traditional and successful approaches to finding carbon stars
in distant galaxies. While the ages of such stars are quite uncertain,
their presence is usually taken as evidence for an intermediate-age
population (Groenewegen 1999). Groenewegen \& de Jong (1993) investigated
the minimum mass for S and C-star formation in the LMC, and derived
1.5 and 1.2~M$_{\odot}$, respectively. The lifetime of stars with these masses
is about 2 and 4.25~Gyr in our Z=0.004 grid. The detection of carbon stars 
therefore suggests that star-formation could have been active in Mrk~178 at ages 
earlier than the 1-2~Gyr frontier. 

The question of the absence or presence of AGB stars has been of great importance
for investigating the SFHs of dwarf elliptical (dE) and dwarf spheroidal (dSph)
galaxies. In such galaxies the argument usually revolves around the detectablility
of low-mass stars in the E-AGB or TP-AGB phase. Low-mass stars, upon ascending the
RGB for the second time, become brighter than the RGB tip in near solar-metallicity
clusters (Frogel \& Elias 1988; Guarnieri, Renzini \& Ortolani 1997). In dwarf
galaxies, with lower metallicity, this may not happen.
Furthermore, in star-forming dwarfs there are so many AGB descendants of
intermediate-mass stars clustering in the same region of the CMD that
it is hard to distinguish the low-mass ones
that would occur just above the TRGB. Age-dating using optical observations
of AGB stars is also made difficult by the strong metallicity dependence of
the isochrones. 
Theoretical models suggest that the near-IR tip-of-the-AGB magnitudes are much less
dependent on metallicity, while retaining a strong age sensitivity
(Bressan, Chiosi $\&$ Fagotto 1994). We can thus use the IR luminosity
function of the stars above the TRGB to bracket the range of ages of
the stellar population in Mrk~178. In Fig. 11 we show the H magnitude 
distribution of the AGB stars, defined as the objects redder than
(J-H)$_o$=0.85 and brighter than the TRGB, having adopted the ``short" distance
modulus of 28.1 so that M$_{H_o,TRGB}$ = --5.5. The luminosity function is populated all
the way up to M$_{H_o}$ = --8.5, which translates into ages of
a few hundred Myr in the Padova isochrones. Stellar generations with a wide range of
ages overlap in this region of the CMD, and in
principle could contribute to the luminosity function.
(We are reluctant to use M$_{H_o}$ as an indicator of M$_{bol}$. 
As discussed in SHGC99, the bolometric corrections are expected to depend on metallicity. If
the reader so wishes, the transformation of Bessell \& Wood (1984) may easily be applied
to Fig.~11, M$_{bol}$ = M$_{H_o}$ + 2.6.) 
If we could unambigously identify the low-mass AGB stars,
then we would have an independent confirmation of star-forming activity taking place
at ages which correspond to look-back times of several Gyr.

Let us now turn to the more luminous AGB stars. The descendents of
intermediate-mass stars.
Gallart et al. (1994) drew attention to the prominence of AGB stars in
modern, CCD-based CMDs of star-forming galaxies. They noticed a very well-populated 
AGB red tail in the Local Group dwarf Irregular (dIrr) NGC~6822. 
According to stellar evolution models, a red tail with extremely red colors 
in [(V-I)$_o$, I$_o$] CMDs occurs only at comparatively high metallicities. 
As the metallicity of the stellar models decreases, 
AGB stars become both bluer and brighter.
Gallart et al. commented on the extremely red tail observed in NGC~6822, as well
as in the LMC and the SMC, and suggested it be used to age-date stellar populations
in such galaxies. An extended red tail appears in the [(V-I)$_o$, I$_o$] CMD
of VII~Zw~403, which presents somewhat of a puzzle (if we take
the isochrones at face value) considering the low metallicity of the ionized gas 
(cf. SHCG99).  As we pointed out repeatedly in our papers on
VII~Zw~403, stellar models are notoriously
uncertain in the AGB phase, but having said that, we proceed with the
comparison because otherwise there is little guidance as to the age of these
stars. In Mrk~178, the red AGB population is less surprising; 
the metallicity of its ionized gas is high enough 
to support the presence of very red AGB stars.   

We proceed to make a purely empirical comparison of the AGB populations in 
different star-forming galaxies. Our color selection criterion includes both E-AGB and
TP-AGB stars. In Mrk~178, we find 56 sources redder than (J-H)$_o$ of 0.85
of the total 791 (or about 7\% of all 
stars detected). In VII~Zw~403, we find 47 of 998 sources (or about 5\% of stars 
detected) in this part of the near-IR CMD. Gallart et al. (1994) noted a 
3\% contribution of AGB stars to the observed, optical CMD of NGC~6822 (about 500 of 16300
sources). The selection effects of this CMD are different from the
near-IR CMDs. Nevertheless the observations do make the point that
modern CMDs are now routinely detecting AGB stars.
Since the lifetimes of the E-AGB and especially the TP-AGB phases 
are short, a large number of progenitors is implied by
the large number of such stars in the CMDs (cf. Renzini 1998).
In other words, the luminous AGB stars require a 
parent population with ages of around a few hundred Myr and above. 
Thus, star-formation in Mrk~178 must have been active at times
of around a few hundred Myr.
 
The near-IR CMD of Mrk~178 is populated by stars with a range
of ages. We find 
young, massive stars in the blue plume; young, intermediate-mass 
AGB and possibly (intermediate-age?) carbon stars in the red plume; and 
older and lower-mass AGB and RGB stars in the red tangle. 
This evidence of star-formation across all ``eras"
of ages which are easily distinguishable with stellar indicators is 
very reminiscent of what we observed for VII~Zw~403. In particular, 
our data do not show any obvious gaps in the stellar distribution 
(meaning, areas of the CMD which could, but are not, populated by stars).
A gap in the distribution of stars is obvious in the CMD of the BCD I~Zw~18. 
Fig.~7 of Aloisi, Tosi \& Greggio (1999) shows a big gap in the red plume,
between the region where one finds the brightest RSGs, and AGB/RGB stars; and it can be 
interpreted with a discontinuous SFH in recent ($0.5$~Gyr) times. 
In other words, I~Zw~18 has undergone at least
two episodes of star-formation separated by a distinct period of quiescence.
The SFHs of Mrk~178 and VII~Zw~403 are different. There are no
such gaps in stellar distribution. These two galaxies are much
closer than I~Zw~18 (by about a factor of two or so); we would have noticed 
gaps. Rather, the CMDs of Mrk~178 and VII~Zw~403 suggest a more-or-less continuous
SFR, which was slightly elevated about 0.5~Gyr ago, and
over a quite extended period of time.

\subsection{Comparison with synthetic CMDs}

In this section we provide synthetic CMDs for a variety of star-forming
histories, and compare them with the Mrk~178 data. The advantage of using
synthetic CMDs over a simple comparison of the data with isochrones or tracks
is that the synthetic CMDs account for the lifetime in different
stellar phases, and thus predict the relative numbers of stars across
the CMD. With the help of synthetic CMDs we can also include,
at least to some extent, the effects of measurement
uncertainties and incompleteness. The synthetic 
CMDs presented here use the aforementioned stellar evolutionary tracks and stellar 
atmospheres, in a simulator presented for the
first time by Tosi et al. (1991) and recently updated for use with HST
WFPC2 data by Greggio et al. (1998). This code has now evolved to simulate NICMOS CMDs. 

Basically, the code is used to calculate CMDs of composite stellar populations 
by random Monte Carlo extractions of stars, following an adopted initial mass 
function (IMF) and SFR law, including errors and incompleteness. 
On the observational side, 
while the errors
and completeness functions were carefully investigated, uncertainties in the
blending properties of stellar images and our distance calibration remain.
The errors inherent to any given synthetic data point depend on how well a particular 
stellar evolution phase is presently understood. 
We note for instance that the simulations do not include
the full TP-AGB phase, as the input tracks terminate at the first pulse. 
Other uncertainties include how well we can transform from the theoretical plane 
(M$_{bol}$, T$_{eff}$) to the observed plane (e.g., M$_{H}$, J-H), 
the treatment of stellar atmospheres, and
transformation to a given filter system. 
In addition, in modeling Mrk~178 it is not clear which metallicites are appropriate for 
the evolving stars. 
Finally, the distribution of errors at a given magnitude in the code does not match 
that for the data.  In other words, the exact distribution of the data, 
including for instance such effects as outlying points outside of
a Gaussian distribution, which will result from blending and near the detection limits, 
is not reproduced by our simulator.
Although this will not affect our main conclusions, future versions 
of the code will need to model better the error distribution.

There are several free parameters that can be varied in order to achieve a match
between the synthetic and the observed CMDs. Our data reach only the very top
of the MS, which is poorly populated, making an IMF determination from
the data impractical. More fundamentally, we should
not derive an IMF from these data, because we are seeing only the
Rayleigh-Jeans tail of their spectral energy distribution. Therefore, all simulations were carried 
out with the Salpeter IMF. The simulator includes stellar masses in the range from
0.6 to 120~M${_\odot}$, so the simulated SFRs and astrated masses refer to this
mass range. It is easy enough to extend the mass range to lower masses, with
any choice of low-mass IMF slope. In many papers on dwarf
galaxies, one finds the use of the Salpeter IMF for the low-mass slope, and
a low-mass cut-off of 0.1~M${_\odot}$. The mass range 
between 0.1 and 0.6~M${_\odot}$ contributes 51\% of the total stellar mass, although
these low-mass stars are ten times as numerous as the more massive ones. 
We apply this correction factor to the total astrated mass. 

Initially, we ran all simulations disregarding the fact that some stars on the CMD  are 
located beyond the first thermal pulse on the tracks.
The Z=0.0004 simulations of Figure~12 are examples of these early simulations.
To account for stars in the TP-AGB phase, we assumed that this phase lasts 1.5x10$^6$~yr 
(Iben \& Renzini 1983) and that they are produced
in the 1.5-3~M${_\odot}$ mass range.  We chose the lower mass limit to 
approximate the minimum mass of carbon stars 
in the LMC.  We first chose an upper limit of 6~M${_\odot}$, to approximate 
the minimum mass of stars which ignite non-violently.  However, 
we found we were getting far too
many stars in the TP-AGB region of the CMD. With an upper mass limit of 3
we still overproduce TP-AGB stars by up to a factor of two to three in some of the simulations.  
This problem is reminiscent of the
missing bright AGB stars in the MCs (e.g. Frogel, Mould \& Blanco
1990). Actually, the observed properties of intermediate age MC clusters and
of field AGB stars now suggest that most of the TP-AGB stars
originate from progenitors less massive than $\approx$3~M${_\odot}$ (e.g.,
Marigo, Girardi \& Chiosi 1996; Maraston 1998). In any case, the final CMDs 
(Fig.~13) show all ``surviving" stars, minus those that would have been in this phase.

In Figure~12, we illustrate a sampling of synthetic CMDs for the two metallicities, Z=0.004 
and Z=0.0004, and the two distance scales, long and short, respectively. 
These CMDs show the surviving stars
of a star-forming event that began 0.1, 2, or 15~Gyr ago and continued to the
present epoch at a constant SFR. The Z=0.004 simulations include the
TP-AGB phase, while the Z=0.0004 simulations were performed with the
simulator before implimentation of the TP-AGB counter. The SFHs adopted 
are quite simple, and provide basic insight into understanding the data. 
The synthetic CMDs also allow us to label each
star according to its evolutionary phase.  We note the several basic
points.

First, Mrk~178 is clearly not a primeval galaxy. The synthetic CMDs which
allow star-formation to occur only in the last 100~Myr are not populated with
any stars in the red tangle. 

Second,  the Z=0.004 grid gives a better description of the
red colors of the luminous red stars than the Z=0.0004 grid does. The low-metallicity
models do not produce E-AGB stars of intermediate mass that are red enough to 
be consistent with
the data, so for ages of up to several hundred Myr, the high-metallicity model
is preferred. Hence, we suppose that
the stellar metallicities of the young stars are indeed similar to that
derived for the gas, at least in the sense that both clearly indicate a 
metallicity higher than Z=0.0004 at the present epoch.

Third, a SFH extending up to 2~Gyr does not yield a distinctive TRGB feature
in the Z=0.0004 simulation. The synthetic CMD for Z=0.004, on the other
hand, does have a feature that could be interpreted as a TRGB. However,
it is at a lower luminosity than the ``observed" TRGB. In this simulation, 
the linear branch of what we had observationally classed as RSGs in section 3, and
the fan-shaped distribution of what we classed AGB stars, are well reproduced.
In either of the two synthetic CMDs stars in the blue plume are overproduced.

Fourth, the synthetic CMDs which extend the SFH to 15~Gyr match better
the small number of stars in the blue plume, and also present a distinct,
TRGB feature. However, the red tangle is too strong, and the
number of luminous red stars above the TRGB is too small. In the Z=0.004 synthetic CMD,
the TRGB is most populated for J-H colors between about 0.9 and 1.2, whereas
the TRGB in the data is most populated for J-H colors $<$0.9. Therefore, the
Z=0.0004 simulation provides for a better description of these data.

Evidently, the very simple SFHs adopted in these first attempts cannot
match all the features of the data. From the previous discussion,  
better models would have a smaller number of stars in the blue 
plume and a distinct TRGB, while maintaining a linear RSG 
feature and a fan-shaped E-AGB/I component.
To do this we constructed a SFH by considering separate regions in the
CMD populated by stars born in different epochs.  We selected
three regions, which we'll briefly call the SG area, at magnitudes
brighter than the TRGB and colors bluer than (J-H)$_o$=0.85; the AGB
area, in the same magnitude region, but at colors between 0.85 and 1;
and the sub-TRGB area, defined as the region fainter than the TRGB.
Comparison with the Z=0.004 tracks (shifted by
m-M=28.8) shows that the
SG area is basically populated only by massive stars,
thus sampling the SF during the last 100 Myr; the AGB area mostly 
ages between 100 Myr and 1 Gyr; the remaining portion of the CMD
samples the epochs older than this.  We ran distinct simulations for
the three regions. The first two were constrained to reproduce the observed number of
objects in the SG and AGB areas, with stars born in the appropriate age
ranges. Each of these simulations brings along synthetic objects
falling in the the sub-TRGB area, according to the stellar evolution
prescriptions. The remaining objects observed in the sub-TRGB area
were then produced by running a third simulation 
with star formation episodes older than 1~Gyr. One model is obviously 
composed of the sum of the three simulations. We notice that we did not 
try to correct for simulated TP-AGB stars, due to the theoretical uncertainty 
affecting their lifetimes and mass range.

We considered two cases for the metallicity: Z=0.004 with the
corresponding distance modulus of 28.8, and a mixed-metallicity case,
in which the RGB stars have Z=0.0004, and the remaining, younger stars
have Z=0.004. Since the distance is derived from the TRGB, the
assumed distance modulus in the mixed-Z simulation is 28.1. With the
short modulus, the evolutionary masses sampled by the SG and AGB areas
are lower than described above, and correspond to different age
bins. In the mixed case we used ages up to 400~Myr, and from 400~Myr to
1~Gyr, for the SG and AGB areas, respectively.
In principle, the entire age range from 1 to 15~Gyr is available for the
third simulation, which had to produce typically 200 stars in the
Z=0.004 case, 350 in the mixed metallicity case. In practice, the
distribution of stars very poorly constrains this age range. We
explored various options for the old SF, on which we comment below.

The numbers of observed stars in the SG and in the AGB areas, which
constrain the relative simulations, are small (a few tens). To model
stastistical effects, we ran simulations in which we allowed the
constraints to vary by the RMS of the number of stars actually present
in the observed CMD.  In Fig.~13 we show two exemplatory synthetic CMDs (with the
oldest episode spanning the whole 1-15~Gyr range), together
with the data. The simulator does not produce the few sparse objects
at the reddest J-H colors. At bright magnitudes, the reddest stars are
TP-AGB objects, counted by the simulator, but not placed on the
CMD. At faint magnitudes the simulator does not reproduce the error
distribution to a great accuracy. A better description of this would
certainly improve the appearance of the synthetic CMDs, but hardly
change our basic conclusions.
 
The salient features of the simulations shown in Fig.~13 are the
following. For the high metallicity simulation, the SFR (corrected to
extend the IMF down to 0.1~M$_\odot$) is 
2.6$\times$10$^{-3}$~M$_\odot$~yr$^{-1}$ in the
last 100~Myr, 5.5$\times$10$^{-3}$~M$_\odot$~yr$^{-1}$ from 1~Gyr to 100 Myr~ago,
and 4.3$\times$10$^{-4}$ M$_\odot$~yr$^{-1}$ from 15 to 1~Gyr ago. This drop comes
about because this is an average SFR over a long period of time. The
set of models of this kind yield a SFR ranging between 
2.6 and 3.7$\times$10$^{-3}$~M$_\odot$~yr$^{-1}$ in the young component; 
between 5.5 and 6.8$\times$10$^{-3}$~M$_\odot$~yr$^{-1}$ in the intermediate 
age component. These
simulations also produce between 159 and 187 TP-AGB stars beyond the
end of our tracks: too many, as stated above.

For the mixed-metallicity simulation, the age bins are 
0 to 400~Myr, 
400~Myr to 1~Gyr, and 1 to 15~Gyr. For the youngest ages, the
SFR derived is between 0.9 and 1.3$\times$10$^{-3}$~M$_\odot$~yr$^{-1}$. These
values are somewhat low compared to the spectroscopically derived rate of
$>$3$\times$10$^{-3}$~M$_\odot$~yr$^{-1}$; however, they also cover
a clearly longer time-interval than the one probed by the H$_\alpha$ emission. (We show
the case of the highest SFR in Fig.~13.)
For the intermediate ages, the SFRs we obtain are between 2.0 and 
7.0$\times$10$^{-3}$~M$_\odot$~yr$^{-1}$. 
(We show our case of 3.5$\times$10$^{-3}$~M$_\odot$~yr$^{-1}$.)
The number of TP-AGB stars produced is between 60 and 173, less of an overproduction
than in the previous set of simulations. For the oldest stars, the average SFR over this
long time interval drops to 4.5$\times$10$^{-4}$~M$_\odot$~yr$^{-1}$.

What can we take away as secure results from these simulations? Fig.~13
indicates that both simulations reproduce well the blue plume and the RSG ``finger"
of the observed CMD. It appears that the SFR of Mrk~178 in the last few 10~Myr
has been of the order of a few times 10$^{-3}$~M$_\odot$~yr$^{-1}$.
This answer comes from both the H$_\alpha$ luminosity, and the simulations
of the supergiant stellar content of the CMD. Both simulations also
produce a stubby, luminous AGB ``finger" with J-H colors redder than 0.85. 
It is here that we expect only a crude qualitative match, since we are missing the TP-AGB phase
in the synthetic CMDs. As we go back to ages of around about
0.5~Gyr, the SFR seems to have been slightly higher. Both sets of simulations give
SFRs that are up by a factor of a few (2 to 5) in this era of the galaxy's history as 
compared to more recent times. For ages upward of 1~Gyr, the SFRs that come out of 
the two sets of simulations are very similar. To compare the simulations with the
data, note that in the data, the red edge of the red tangle is at a J-H
color of 0.9. Beyond a color of 0.9, the stellar numbers drop off; it is here that
we also get the data with the highest error bars, which are not captured well by the simulations.
The simulation with Z=0.004 produces a red tangle which is clearly too red.
It has the morphological appearance of a third, RGB ``finger", which sticks out
of the red tangle at a slant. 
The Z=0.0004 red tangle does not exhibit this morphology. It reproduces well
the overall, triangular shape of the red tangle in the data. However, the
red edge in this red tangle is too blue by about 0.1 mag, giving the appearance
of too many stars in the red tangle. We have a slight preference
for a predominantly low-metallicity RGB, because it yields a red tangle
that is located beneath the luminous portion of the red plume, a
``sharp" red edge for the red tangle, and an overall
distribution of stars near the TRGB which reproduces
the data quite well. 

We now describe our efforts to set some limits on the SFH of Mrk~178 beyond the age of
1~Gyr. Our first argument is based on the total mass in stars. 
We derived the SFR over a certain time interval, so multiplying the SFR with
this time interval yields the total astrated mass encompassed by the NIC2 field. 
Since the NIC2 field was located 
near the center of Mrk~178, let us assume the mass we derive is
representative of a good fraction of the galaxy's mass. Stars that populate 
the red tangle, in particular those in the oldest of our age bins,
strongly influence our result concerning the early star-formation in Mrk~178 
and thus, its total astrated mass.  
It is possible to find good matches of 
the observed CMD for a variety of fractional contributions of 
intermediate versus low-mass stars to the red tangle. 
If more of the red tangle is thought to arise from higher mass stars, 
the total astrated 
mass is lower. The lowest total astrated masses consistent with
our data are from models which constrict the SF to the 1 to 2~Gyr age range.

For example, if we add in the high-metallicity simulation the 200 stars in
the RGB area over the age interval of 1 to 15~Gyr (as was done in the
exemplatory simulation discussed above), they are drawn from
a population with a total mass of about 6x10$^6$~M${_\odot}$. The same is
true if we add over this time interval the 350 RGB stars of low-metallicity
(since the average SFRs of the two cases simulated are so similar). If instead
we add them only over an age range of 1 to 2~Gyr, then the total mass
drops down to about 3x10$^6$~M${_\odot}$. Finally, if for instance
we add these stars between
14 and 15~Gyr, then the mass goes up to just over 1x10$^7$~M${_\odot}$.
Thus, while in general both the time and duration of low-mass star
formation control the mass of the galaxy, we find
that a short event early on in the history of the galaxy produces a higher
mass than long-term, low-level star-forming activity, and a short event late
in the history of the galaxy. 

We have little guidance concerning
the amount of mass we really should expect to be locked up in stars. 
Observed masses for BCDs usually are for the neutral gas (H-I) content, or some
dynamical mass that includes the dark matter. The blue luminosity derived
in section 3.2 cannot be
used, as it is dominated by the young stars with a very different mass-to-light
ratio than that of the old stars. Nonetheless, note that the highest
astrated mass can be achieved by adding most stars early in the galaxy's life, and
that the total mass in stars can in this way be raised to be of the order of
10$^7$~M${_\odot}$. This is of the same order of magnitude as the H-I mass of
Mrk~178. If, on the other hand, we assume that this galaxy is no older
than 2~Gyr, in other words, if we assume all of the stars were added
just at the 1-2~Gyr frontier, then the total astrated mass is only a few times
10$^6$~M${_\odot}$. While the differences are not dramatic (a factor of three or so), 
one wonders whether the stellar mass of a BCD is comparable to the mass of a very
massive globular cluster like $\omega$~Cen (Pryor \& Meylan 1993), or whether
it should instead be significantly higher. Unfortunately we cannot be sure how much
mass is contained in the background sheet discussed in the next section. As Fig.~1
indicates, the NIC2 field was located near the bright central regions of the elliptical
light distribution, which we assume to dominate the galaxy's mass. If the stars in the
background sheet contribute a few (as in, up to 10) times the mass encompassed by the NIC2 field, 
then our guesses for the stellar mass independently of the exact formation age are 
of the order of 10$^7$-10$^8$~M${_\odot}$.

The determination of galaxy masses is one of the potential
strenghts of CMD studies, but to accomplish this, the CMDs need to be deep enough 
to reveal directly the low-mass stars. Once the SFR at all ages is determined,
it can in principle be integrated to derive a value for the total mass. This is
already being done for Local Group dwarfs. For the more distant BCDs,
this will perhaps become feasible observationally for a few nearby examples with the NGST.
Suffice it
to say that the hypothesis of Izotov \& Thuan, in which BCDs are either primeval
for extremely low metallicities, or no older than 1-2~Gyr in the case of low metallicities,
goes hand in hand with saying that these galaxies have low astrated masses.

Our second argument addresses the question whether or not Mrk~178 could be
a faint blue galaxy. Following the hypothesis of
Babul \& Ferguson (1996), the faint blue galaxies are small, low-mass galaxies that
undergo very short (10$^7$~yr) starbursts at redshifts of about 0.5 to 1.
The number of stars in the red tangle now is used to constrain the burst strength.
The SFRs in the simulations of the 1-2~Gyr or 14-15~Gyr events in Mrk~178 are 
not high, of the order of several times 10$^{-3}$M${_\odot}$yr$^{-1}$. Therefore, a Gyr-long
SF event would not be able to generate the SFR needed in the Babul \& Ferguson 
model; it falls short by several orders of magnitude.
Even a short (0.1~Gyr) burst at intermediate (5 or 7.5~Gyr) age can bring
the SFR up to only a few times 10$^{-2}$M${_\odot}$yr$^{-1}$, and still falls short
of the 1-10~M${_\odot}$yr$^{-1}$ assumed in the Babul \& Ferguson scenario. An 
extremely short, 10~Myr starburst, between 7.5 and 15~Gyr, matching pretty closely the 
input parameters of the Babul \& Ferguson models, raises the SFR to 
0.6-1~M${_\odot}$yr$^{-1}$; this is approaching the order-of-magnitude SFR required.
In other words, it is {\it possible} that the red tangle of Mrk~178 hides an event that
is consistent with that required to call it a faint blue galaxy. However, to accomplish
this the galaxy had to form over an extremely short time period and at early times, as well as 
remain inactive until the onset of the recent SF activity.

In summary, the CMD simulations have given us the following insights into the SFH of
Mrk~178. As in VII~Zw~403 (cf. Lynds et al. 1998), the SFR seems to have been 
higher about a few hundred Myr ago, then it was in the more recent past. 
The CMD requires
that stars formed more than 1~Gyr ago --- and the older the stars, the better defined
the TRGB and the higher the astrated mass. A very short burst of
star formation at an early epoch can be hidden in the data; 
Mrk~178 could be a descendent of the Faint Blue Excess population of galaxies in deep images.

\subsection{Morphology}
 
Loose \& Thuan (1986) performed a deep CCD imaging survey of 50 BCDs from the list
of Thuan \& Martin (1981). They found that over 95\% of BCDs show red halos of
light which extend beyond the star-forming regions and which suggest the
presence of an old, underlying stellar population. These galaxies
are the best candiates among the BCDs for harboring dynamically relaxed, old 
stellar substrata. Conversely, the question of whether or not the remaining few 
objects for which such regular, extended outer isophotes have not been discovered 
are primeval galaxies remains a matter of great interest  
(e.g. Izotov \& Thuan 1999, SHCG99).

In the case of VII~Zw~403, we were able to demonstrate that the RGB population not
only underlies the centrally located star-forming region, but is the dominant
population at large galacticentric radii. Previously,
the red color derived from spatially integrated surface photometry of the halos of BCDs
was used to infer the presence of an ancient background sheet of stars (e.g.
Loose \& Thuan 1986, Kunth, Maurogordato \& Vigroux 1988, Papaderos et al. 1996, 
Telles, Melnick \& Terlevich 1997, Meurer 1999). However, RGB, AGB and RSG stars overlap
in temperature, and hence color. The direct
detection of individual RGB stars far from the center of VII~Zw~403 (SHCG99)
provided the necessary evidence that the integrated red colors may indeed be due
to RGB stars and thus a stellar population that is significantly older than the starburst. 

For Mrk~178, we detect an RGB substratum in the inner regions of the galaxy, but we have
no resolved stellar photometry for the halo. The DSS image (Fig.~1) and available CCD 
images, however, testify to the presence of extended outer isophotes with regular, 
elliptical contours. To emphasize this point, we show in Fig.~14 a slightly manipulated 
version of the DSS image. We performed a sky subtraction and smoothing with a 3-pixel 
($\approx$5") boxcar filter, and show the result as a contour plot. From this plot we infer that 
the major axis may be traced to at least 2', 
much further out than the size listed in the RC3 (1\farcm23 to the standard, 25-magnitude level).
This illustrates that indeed a very faint and very extended background sheet exists.
The stars populating the extended halo contribute to the total mass of the galaxy,
but with the data in hand, we cannot derive a reasonable estimate of how much.

There are no surface-brightness profiles available for Mrk~178. Huchra (1977) gives 
aperture photometry in a series of increasing circular apertures. In Table~1, 
we reproduce colors in the smallest aperture,
which happens to coincide closely with the size of the NIC2 field, and in two rings
of increasing size.  
According to Table~1, the colors are bluest,
in the smallest, 24" measurement aperture, and rise to an intermediate color in
the 24"-56" ring and the 38"-56" ring. 
The data thus suggest that Mrk~178 becomes redder 
outside of the star-forming core. This is the behavior generally observed
for type iE BCDs. 

We compared the aperture photometry of Mrk~178 with the integrated, empirical star-cluster
colors of Schmidt, Alloin \& Bica (1995). Schmidt, Alloin \& Bica derived these data 
using as spectral templates young and intermediate-age clusters in the SMC, LMC and Galaxy disk. 
Their dwarf Elliptical/dwarf Spheroidal (dES) galaxy template is 
constructed using spectral libraries of three metal-poor galactic globular clusters. 
No U-V colors are available
in the Schmidt, Alloin \& Bica database. The observed colors of Mrk~178 were overlayed onto
their Fig.~9 diagnostic diagram. It is evident that the color in the smallest 
aperture is consistent 
with that of the youngest cluster templates in the age range from 20~Myr to 200~Myr. 
The colors in the rings, on the other hand, can be matched
by clusters with ages of 200 to 700~Myr.  

We interpret the color at radii beyond the NIC2 field as follows.
Young stars are the dominant population in the core where the NIC2 field
was located; they outshine any older stars in integrated light.
The young, blue stars are concentrated near the center, in the two
prominent H-II regions. This component to the integrated light decreases
 away from the core of the galaxy toward its halo, so that 
the older stars become a more important contribution to the
integrated light at larger radii.
This is reminiscent of the core-halo structure
seen in the resolved stars in VII~Zw~403 (SHCG99).

To summarize, Mrk~178 exhibits the same iE morphology as VII~Zw~403, the  
most common type in the Loose \& Thuan sample. 
We have now demonstrated the existence of RGB stars in two such BCDs. 
This supports the hypothesis put forth based on the morphology and the colors of
the red halos of BCDs (Loose \& Thuan 1986, Kunth, Maurogordato \& Vigroux 1988, 
Papaderos et al. 1996, Telles, Melnick \& Terlevich 1997, Meurer 1999) that 
all BCDs with extended, red background sheets harbor old stars.

\section{Conclusions}

Mrk~178 is a typical BCD in terms of its morphology, size, luminosity, H-I mass, and
gas abundances. In its H-II regions, its star-formation rate is at least a few times 
10$^{-3}$M${_\odot}$yr$^{-1}$. This is typical for the SFR of a late-type dwarf.

We use near-IR photometry with the HST to find the
RGB, showing that Mrk~178 must be at 
least 1-2~Gyr old. The older this population the better it fits at the TRGB, and
the higher the total astrated mass of the galaxy. Employing a calibration previously derived by us
for the TRGB method in the near-IR, we determine a minimum distance of 4.2($\pm$~0.5)~Mpc
for Mrk~178. This distance is 17$\%$ larger than that based on ground-based photometry
of the brightest resolved objects.  
In applying the near-IR TRGB method for the first time to a galaxy for which there
is no optical TRGB, we encountered several problems. First, lacking a metallicity calibration 
for the RGB in the near-IR, the distance estimate assumes that
--2.3$<$ [Fe/H] $<$--1.5 for old RGB stars. A higher metallicity has the effect of 
increasing the distance. Second, Mrk~178 exhibits a very strong AGB component. These
stars have similar colors as RGB stars and luminosities that extend above the TRGB. Their effect
is to add noise to the luminosity functions from which we are attempting to determine the 
onset of the TRGB feature. This is an obvious drawback of the near-IR TRGB method.

Luminous AGB stars account for as much as $7\%$ of the resolved stars in our CMD, and populate
a distinct area in color and luminosity space. Their presence indicates
that star-formation was active and strong several hundred Myr ago. 
The upper limit to the age of the stars in the AGB area brighter than the TRGB is difficult
to establish, since tracks of a wide range of initial masses overlap
in this region. There may even be a contribution from stars as old as
several Gyr, as inferred for early-type dwarfs.  

We find several objects with extremely red near-IR colors and suggest that they are
carbon stars. The detection of candidate carbon stars in BCDs is a result which
only became possible with the advent of deep near-IR photometry. 

We use Monte-Carlo simulations to constrain the SFH quantitatively. 
For stellar ages below 1~Gyr, the observed CMD provides good 
constraints on the SFH. 
The prominent red-tangle feature potentially contains truly ancient stars.
Synthetic CMDs with a range of SFHs all produce reasonably matching 
red tangles. 
However, when the red tangle is attributed exclusively to relatively young stars with
ages between 1 and 2~Gyr, then
the total astrated mass of the galaxy (within the NICMOS field-of-view) is as small as that of
a massive globular cluster. The further the SF extends to earlier ages,
the higher the total astrated mass. 

In order to explore whether or not Mrk~178 could be the descendant of a faint blue galaxy,
we modeled the SFH beyond 1~Gyr with  short bursts. The total number of available stellar
progeny of an intermediate-age burst in the red tangle limits the possible burst strength. 
Assuming a burst duration of 10~Myr, such a potential burst was of the right
order of magnitude in strength compared to what is needed 
in the Babul $\&$ Ferguson (1996) scenario.

The main question we set out to answer was whether Mrk~178 is young or old.
While our CMDs alone cannot prove the existence of RGB stars 
older than a few Gyr, we argue that
the detection of RGB stars in combination with the morphology of Mrk~178 provides a
strong indication for the presence of low-mass stars and hence, a large formation age
of this galaxy.

\acknowledgments Work on this project was supported through HST grants to RSL and 
MMC (projects 7859 and 8012). UH acknowledges financial support from SFB~375. We thank
Quentin Bailey for writing a piece of code that allowed RSL to compare luminosity
functions. We also thank Dr. Deidre Hunter for sharing her unpublished data.
We made extensive use of the SIMBAD and NED data bases.

\bigskip

NOTE ADDED IN PROOF: After this paper was accepted, we learned that
in February of 2000, P.~Papaderos of the University of G\"{o}ttingen obtained deep, 
ground-based CCD images of Mrk~178 at the Calar Alto 1.23-m telescope.
He communicates to us that his images in the Johnson-Cousins B and R
system clearly exhibit an extended red sheet which is very faint,
regular, and elliptical in shape; he also identifies about 20 
faint star-forming regions in the north-eastern 
parts of the galaxy which were not encompassed by the NIC2 FOV. 
The color profile shows that outside of 30---35$\arcsec$ (well outside of 
the central, dominant H-II regions at which we pointed the NIC2) 
Mrk~178 becomes quite red, (B-R)$_0$ $\approx$ (1.1$\pm$0.1)~mag, 
supporting the idea that this is not a young galaxy.

\clearpage

\figcaption[] {The 4'x4' 103a-E image of Mrk~178 from the Digitized Sky Survey
in grey-scale representation. The brightest feature is a blend of the two main
H-II complexes. This is where the NIC2 camera was pointed. The faint, elliptical 
main body can also be seen. 
}

\figcaption[]{NIC2 color image composed from combining the images taken
through the F110W and F160W filters. This image is about 19"x19". }

\figcaption[]{The DAOPHOT-derived photometric errors.}

\figcaption[]{Completeness fractions from ADDSTAR tests.}

\figcaption[]{Color-magnitude diagrams of Mrk~178 in magnitudes
in the HST Vega system. The location of the TRGB is indicated.
The luminosity function in F110W (see Fig.~6) exhibits a sharp jump; this is
where we mark the TRGB in the upper panel. The luminosity function in
F160W displays a gradual rise; this is due to the slanted TRGB in
the bottom panel (where our estimated errors are indicated by dashed
red lines). }

\figcaption[]{Luminosity functions of Mrk~178 in magnitudes in the HST Vega system
for stars with 0.75~$<$ (F110W-F160W) $<$~1.5. The location
of the TRGB is indicated. Notice the high completeness at the TRGB. We can
also see the contribution to the luminosity functions by AGB stars 
at magnitudes brighter than the TRGB.}

\figcaption[]{Color-magnitude diagrams of Mrk~178 after transformation
to ground-based J and H. The foreground extinction is negligible. }

\figcaption[]{Color-magnitude diagrams of Mrk~178 with three cuts applied on the
size of the photometric errors (1${\sigma}$) in both J and H.
The salient features of Fig.~7 remain when only data with good errors are
being considered. Notice by comparison with Fig.~7 or 9 that many 
faint and very red, and some faint and blue stars, have large errors.
}

\figcaption[]{Color-magnitude diagrams of Mrk~178 with two
sets of tracks overlayed. The first set of tracks is for a metallicity
of Z=0.0004 and the corresponding short distance modulus of 28.1. The
second set is for Z=0.004 and the long distance modulus of 28.8. 
For each case, the TRGB is indicated, and the corresponding 
absolute-magnitude scale is given on the right-hand ordinate. }

\figcaption[]{A comparison of the near-IR CMDs of VII~Zw~403 and Mrk~178. 
The data for Mrk~178 are selected to have J and H errors smaller than
0.2~mag. The color-coding
is blue for MS, BSG, and BL stars; magenta for RSG and luminous AGB stars; red
for RGB and faint AGB stars; black for AGB stars;
with black circles (in the case of VII~Zw~403) and large black dots
(in the case of Mrk~178) for near-IR identified AGB stars 
(presumably in the TP-AGB phase). 
For more details, refer to the text. The TRGB is shown
by dashed green lines. The dashed blue lines indicates the location of the
blue plume. The dashed red line indicates where more stars might belong
to the RGB phase in Mrk~178 and the black dashed line indicates where
optically-identified AGB stars are separable from RSGs in the VII~Zw~403 CMD.}

\figcaption[]{The number of AGB stars detected in the near-IR as a function of 
their luminosity in Mrk~178 and VII~Zw~403. Stars just above the TRGB, at around
M$_H$=--6, could be low-mass and hence old, AGB stars (see text).}

\figcaption[]{Synthetic CMDs for Z=0.0004 (left) and Z=0.004 (right),
using the Salpeter IMF, the appropriate distance, and a constant SFR, in the interval
from the maximum time indicated in each panel to the present epoch. The simulator
for Z=0.0004 did not have the TP-AGB counter enabled, so all survivors are
plotted. The Z=0.004 simulator shows all survivors minus the TP-AGB stars,
for which we do not have tracks or atmospheres.
The dashed lines drawn indicate the observed location of the TRGB, the center
of the blue plume, and the ``dividing line" of the RSG vs. (optically detected) 
AGB stars.}

\figcaption[]{A comparison of the data with two exemplary, synthetic CMDs.
For these complex SFHs, the history of the galaxy was divided into 3 age bins. 
The simulation labeled ``Z=0.004" uses high-metallicity models and assumes
the long
distance scale. The simulation labeled ``Z mixed" uses Z=0.0004 for stars with 
ages greater than 1~Gyr and Z=0.004 for younger stars, and assumes the short distance scale.} 

\figcaption[]{A contour-plot of the DSS image, after sky subtraction and
smoothing with a 3-pixel ($\approx$5") boxcar filter.
This representation shows very well the large extent of the faint, elliptical halo, and 
emphasizes the type ``iE" morphology of Mrk~178.} 

\end{document}